\documentclass[aps,prd,twocolumn,showpacs,preprintnumbers,nofootinbib,amsmath,amssymb]{revtex4}

\usepackage{graphicx}% Include figure files
\usepackage{dcolumn}% Align table columns on decimal point
\usepackage{bm}% bold math
\usepackage{amsmath}
\usepackage[normalem]{ulem}

\newcommand{\beq}{\begin{eqnarray}}
\newcommand{\eeq}{\end{eqnarray}}

\def\spose#1{\hbox to 0pt{#1\hss}}
\def\ltapprox{\mathrel{\spose{\lower 3pt\hbox{$\mathchar"218$}}
 \raise 2.0pt\hbox{$\mathchar"13C$}}}

\begin{document}

\title{
Anisotropy of the quark-antiquark potential in a 
magnetic field
}
\author{Claudio Bonati}
\email{bonati@df.unipi.it}
\affiliation{
Dipartimento di Fisica dell'Universit\`a
di Pisa and INFN - Sezione di Pisa,\\ Largo Pontecorvo 3, I-56127 Pisa, Italy}

\author{Massimo D'Elia}
\email{delia@df.unipi.it}
\affiliation{
Dipartimento di Fisica dell'Universit\`a
di Pisa and INFN - Sezione di Pisa,\\ Largo Pontecorvo 3, I-56127 Pisa, Italy}

\author{Marco Mariti}
\email{mariti@df.unipi.it}
\affiliation{
Dipartimento di Fisica dell'Universit\`a
di Pisa and INFN - Sezione di Pisa,\\ Largo Pontecorvo 3, I-56127 Pisa, Italy}

\author{Michele Mesiti}
\email{mesiti@df.unipi.it}
\affiliation{
Dipartimento di Fisica dell'Universit\`a
di Pisa and INFN - Sezione di Pisa,\\ Largo Pontecorvo 3, I-56127 Pisa, Italy}

\author{Francesco Negro}
\email{fnegro@pi.infn.it}
\affiliation{
Dipartimento di Fisica dell'Universit\`a
di Pisa and INFN - Sezione di Pisa,\\ Largo Pontecorvo 3, I-56127 Pisa, Italy}

\author{Francesco Sanfilippo}
\email{f.sanfilippo@soton.ac.uk}
\affiliation{School of Physics and Astronomy, University of Southampton, SO17 1BJ Southampton, 
United Kindgdom}

\date{\today}% It is always \today, today,
             %  but any date may be explicitly specified

\begin{abstract}
We investigate the static 
$\overline{Q}Q$-potential for  $N_f = 2+1$ 
QCD at the physical point 
in the presence of a constant and uniform 
external magnetic field.
The potential is found to be anisotropic
and steeper in the directions transverse to the magnetic
field than in the longitudinal one.
In particular, when compared to the standard case with zero background field,
the string tension
increases (decreases) in the transverse (longitudinal) direction,
while the absolute value of the Coulomb coupling and the Sommer parameter
show an opposite behavior. 
\end{abstract}

\pacs{12.38.Aw, 11.15.Ha,12.38.Gc,12.38.Mh}
%\keywords{Suggested keywords}%Use showkeys class option if keyword
                              %display desired
\maketitle

\section{Introduction}
\label{intro}

The properties of strong interactions in the presence of 
a strong magnetic background have attracted much 
interest in the last few years
(see, e.g., Ref.~\cite{lecnotmag} for recent 
reviews). This is justified by the 
many contexts in which such properties may play a role:
the physics of 
compact astrophysical
objects~\cite{magnetars}, 
of non-central heavy ion collisions~\cite{hi1,hi2,hi3,hi4, tuchin} 
and of the early Universe~\cite{vacha,grarub}, 
involves fields going from $10^{10}$ Tesla 
up to $10^{16}$ 
Tesla ($|e| B \sim 1$ GeV$^2$).

One important feature
is that gluon fields, even if not directly coupled to electromagnetic
fields, may undergo significant modifications, through 
effective QED-QCD interactions induced by quark loop 
effects~\cite{anisotropic, chernodub, musak, elze1, elze2, mueller, galilo, KojoSu1, 
KojoSu2, watson, andersen, ozaki, kamikado, mueller2, simonov2}.
Such a possibility has been confirmed
by lattice QCD 
simulations~\cite{demusa,DEN,reg0,Ilgenfritz:2012fw,reg2,kovacs,EB,Ilgenfritz:2013ara}, 
resulting also 
in unexpected behaviors, like inverse magnetic catalysis~\cite{reg0, 
catalreview, kovacs, fukuhida, Ilgenfritz:2013ara, Bornyakov:2013eya, Chao:2013qpa, 
Fraga:2013ova, Yu:2014sla, Ferreira:2014kpa, Farias:2014eca, Ruggieri:2014bqa}.

One of the main attributes of strong interactions is the appearance of 
a confining potential. In the heavy quark limit,
that is related to the expectation value of Wilson loops, 
so that the potential emerges as a
property of gauge fields only. It is interesting, then, to ask 
whether a magnetic background can influence the 
static quark-antiquark potential.
Many studies have considered the possible emergence
of anisotropies~\cite{anisotropic, chernodub, galilo, ozaki}, 
which may have many phenomenological implications, like a modification
of the heavy quark bound states.

On the lattice, anisotropies in the gauge field distributions
have already been observed in quantities like the 
average plaquettes 
taken in different planes~\cite{Ilgenfritz:2012fw, reg2}. 
However, an investigation
of the possible anisotropies 
of the static potential  
is still missing.

In this paper, we present
an exploratory study of this issue,
based on numerical simulations
of $N_f = 2+1$ QCD with physical quark masses,
discretized by
stout improved staggered fermions and the tree level improved 
Symanzik gauge action. 
In particular, we look at the 
expectation values of Wilson loops in the 
presence of a magnetic background, and  
compute from them the static potential
for quark-antiquark separations orthogonal 
or parallel to the magnetic field. 
We consider zero 
temperature, three different lattice spacings 
and magnetic fields up to  
$|e|B \sim 1$ GeV$^2$. Results show the emergence 
of anisotropies both in the string tension and in the
Coulomb part of the potential.

\section{Numerical Methods}

We have considered 
a constant and 
uniform magnetic field, which enters the QCD Lagrangian through quark 
covariant derivatives $D_\mu = \partial_\mu+igA_\mu^aT^a+iq_fA_\mu$,
where $A_\mu$ is the Abelian gauge potential and $q_f$ is the quark 
electric charge. On the lattice, that amounts 
to adding proper $U(1)$ phases to the usual $SU(3)$ links entering
the Dirac operator. The Euclidean partition function is expressed as
\begin{eqnarray}
\label{partfunc}
\mathcal{Z}(B) &=& \int \!\mathcal{D}U \,e^{-\mathcal{S}_{Y\!M}} \!\!\!\!\prod_{f=u,\,d,\,s} \!\!\! \det{({D^{f}_{\textnormal{st}}[B]})^{1/4}}, \\
\label{tlsyact}
\mathcal{S}_{Y\!M}&=& - \frac{\beta}{3}\sum_{i, \mu \neq \nu} \left( \frac{5}{6}
W^{1\!\times \! 1}_{i;\,\mu\nu} -
 \frac{1}{12} W^{1\!\times \! 2}_{i;\,\mu\nu} \right), \\
\label{fermmatrix}
(D^f_{\textnormal{st}})_{i,\,j}&=&am_f \delta_{i,\,j}+\!\!\sum_{\nu=1}^{4}\frac{\eta_{i;\,\nu}}{2} 
\left(u^f_{i;\,\nu}U^{(2)}_{i;\,\nu}\delta_{i,j-\hat{\nu}} \right. \nonumber\\
&-&\left. u^{f*}_{i-\hat\nu;\,\nu}U^{(2)\dagger}_{i-\hat\nu;\,\nu}\delta_{i,j+\hat\nu}  \right)
\end{eqnarray}
where $\mathcal{D}U$ is the functional integration over the $SU(3)$ 
link variables. 
$\mathcal{S}_{Y\!M}$
is the tree level improved Symanzik action~\cite{weisz,curci},
involving the real part of the trace of the 
$1\!\times \! 1$ 
($W^{1\!\times \! 1}_{i;\,\mu\nu}$)
and $1\!\times \!2$ 
($W^{1\!\times \! 2}_{i;\,\mu\nu}$) 
loops.
$D^f_{\textnormal{st}}$ is the staggered Dirac operator, built up by
two times stout-smeared links $U^{(2)}_{i;\,\mu}$~\cite{morning}
(with isotropic smearing parameters $\rho_{\mu\nu}=0.15\,\delta_{\mu\nu}$),
in order to reduce finite cut-off effects, in particular
taste symmetry violations {(see \cite{bazavov} for a comparison 
of the effectiveness in reducing taste violations of the different 
improved staggered discretizations)}.

The Abelian continuum gauge field $A_y=B x$ and $A_\mu=0$ for $\mu=t,\,x,\,z$, 
corresponding to a magnetic field in the $\hat z$ direction,
is discretized on the lattice torus as 
\begin{eqnarray}
\label{bfieldy}
&& u^f_{i;\,y}=e^{i a^2 q_f B i_x} \ , \ \ \ \ \ \ 
{u^f_{i;\,x}|}_{i_x=L_x}=e^{-ia^2 q_f L_x B i_y} \ \ \ \ 
\end{eqnarray}
with $u^f_{i;\,\mu}=1$ elsewhere and $B$ is quantized as~\cite{thooft, bound3, wiese, review}
\beq
|e| B = {6 \pi b}/{(a^2 L_x L_y)} \, ; \ \ \ \ \  b \in \mathbb{Z} \, .
\label{bquant}
\eeq
No stout smearing is applied to the $U(1)$ phases, which 
are treated as purely external parameters (quenched QED approximation).

\begin{table}[b]
\centering
\begin{tabular}{ |c|c|c|c|c|c| }
\hline
$L$ & $a (\textnormal{fm})$ & $\beta$  & $am_{u/d}$ &$am_s$ & b \\
\hline
24 & 0.2173(4) & 3.55  &  0.003636 & 0.1020 & 0,12,16,24,32,40\\
32 & 0.1535(3) & 3.67  &  0.002270 & 0.0639 & 0,12,16,24,32,40\\
40 & 0.1249(3) & 3.75  &  0.001787 & 0.0503 & 0,8,12,16,24,32,40\\
\hline
\end{tabular}
\caption{Simulation parameters, 
chosen according to Refs.~\cite{tcwup1,befjkkrs} and 
corresponding to a physical 
pion mass. The systematic error on $a$ 
is about $2\%$~\cite{tcwup1}.}
\label{param}
\end{table}

We performed simulations at the physical value of the pion mass, 
$m_\pi \sim 135$ MeV, and three different values of 
the lattice spacing $a$,
using the bare parameters 
reported in 
Table~\ref{param} 
($m_s/m_{u,d}$ is fixed to its physical value, 28.15).
{In Ref.~\cite{tcwup1} it was shown that making use of different
standard observables (like for instance $m_K$ or $f_K$) to 
determine the physical point, compatible results are obtained,
so we do not expect large cut-off effects to be present.}
We explored symmetric, zero temperature lattices, 
with the number of sites per direction ($L$)
tuned to maintain $L a \simeq 5$ fm.
The Rational Hybrid Monte-Carlo (RHMC)
algorithm was used to sample gauge 
configurations,
with statistics ranging, for each value of 
$B$, from $O(10^4)$ 
to $O(10^3)$ 
molecular dynamics (MD)  
time units,
going from the coarsest to the finest 
lattice.

In order to determine the spin-averaged potential between a static
$\overline{Q}Q$ pair, separated by a distance $\vec R$, we considered
the large time behavior of the average rectangular 
Wilson loop $W(\vec R,T)$, where $T$ is the time extension of 
the loop. Usually, based on space-time
isotropy, one averages over all
directions of $\vec R$; on the contrary, apart from the 
$B = 0$ case, we considered   
separately the averages over different directions,
thus leaving room to the possibility
that $V(\vec R)$ may not be a central potential.

Going to a lattice notation, in which
$\vec n $ and $n_t$ denote the dimensionless spatial 
and temporal sides of the loop, the potential
can be obtained as
\begin{equation}
aV(a \vec n)= \lim_{n_t\to\infty} 
\log \left(\frac{\langle W (a \vec n, a n_t)\rangle}
{\langle W (a \vec n, a(n_t+1))\rangle}\right) \, .
\label{combination}
\end{equation}
In practice, one plots the right-hand side as a 
function of $n_t$ and looks for a stable plateau at large times, 
from which the potential can be extracted by a fit to a 
constant function. In the present study we limit ourselves 
to the  cases of $\vec n$ parallel or orthogonal 
to $\vec B$.

For each simulation, we measured Wilson loops every 
5 MD time units. 
In order to reduce the UV noise, we applied 
one step of
HYP smearing~\cite{Hasenfratz:2001hp} for 
temporal links, with smearing parameters 
corresponding to the so-called HYP2-action of 
Ref.~\cite{Della Morte:2005yc}, 
 and $N_{SM}$ steps of 
APE-smearing~\cite{Albanese:1987ds} for spatial links, with
smearing parameter $\alpha_{\rm APE} = 0.25$.

Since APE-smearing treats all spatial directions symmetrically, it is important
to check that possible anisotropies be not washed out by this 
noise reduction technique. We studied, for a few cases, the dependence of results on the number of 
smearing steps and, 
having checked that it is not significant (see next section),
we fixed $N_{SM} = 24$.
The statistical errors on the right-hand side of Eq.~(\ref{combination}),
as well as those on the parameters of the fitted plateaux,
were determined by performing a bootstrap analysis to
take correlations 
into account.

\begin{figure}[t!]
\includegraphics[width=0.92\columnwidth, clip]{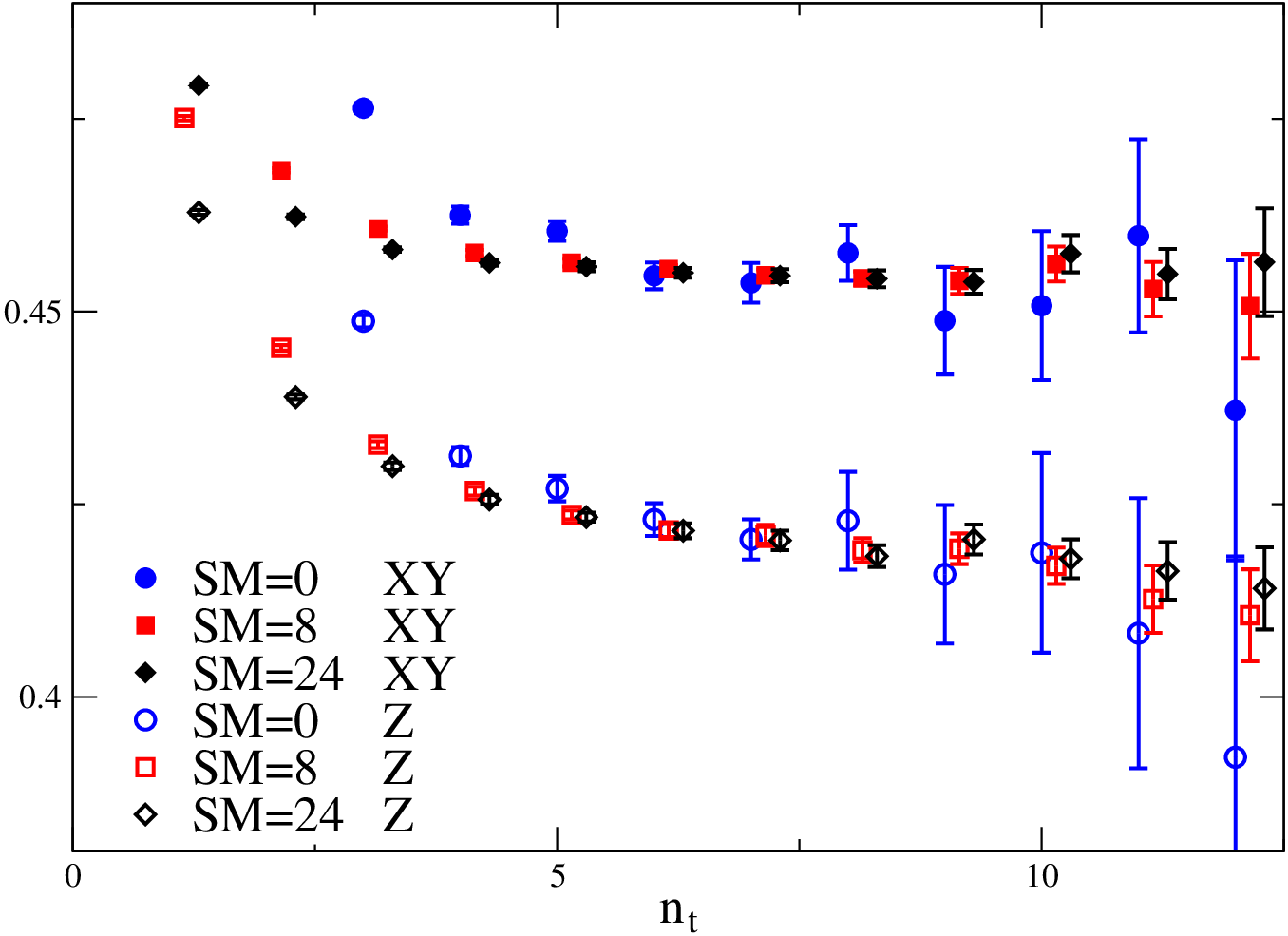}
\caption{Wilson loop combination defined in Eq.~(\ref{combination}) for $|\vec{n}|=3$ as a function of $n_t$ and for several 
values of the APE smearing level. The simulation was performed on the $32^4$ lattice at 
$|e|B=0.97\ \textnormal{GeV}^2$. }
\label{fig_plateau}
\end{figure}

\section{Numerical Results}

In Fig.~\ref{fig_plateau} we report an example of
the logarithm of Wilson loop ratios, see Eq.~(\ref{combination}), as a function
of $n_t$ and for different APE-smearing levels, 
obtained at spatial distance $|\vec{n}|=3$ and 
for $|e|B=0.97\ \mathrm{GeV}^2$. We show separately results averaged over the 
longitudinal ($Z$) or transverse ($XY$) directions.
A well defined plateau is visible in both cases, and the emergence
of anisotropies clearly appears, with the potential being
larger in the transverse direction. It is also evident that 
smearing has no visible effect on such anisotropies, 
so that one 
can safely adopt a number of smearing levels 
large enough to 
have a good noise/signal ratio.

\begin{figure}[t!]
\includegraphics*[width=0.95\columnwidth, clip]{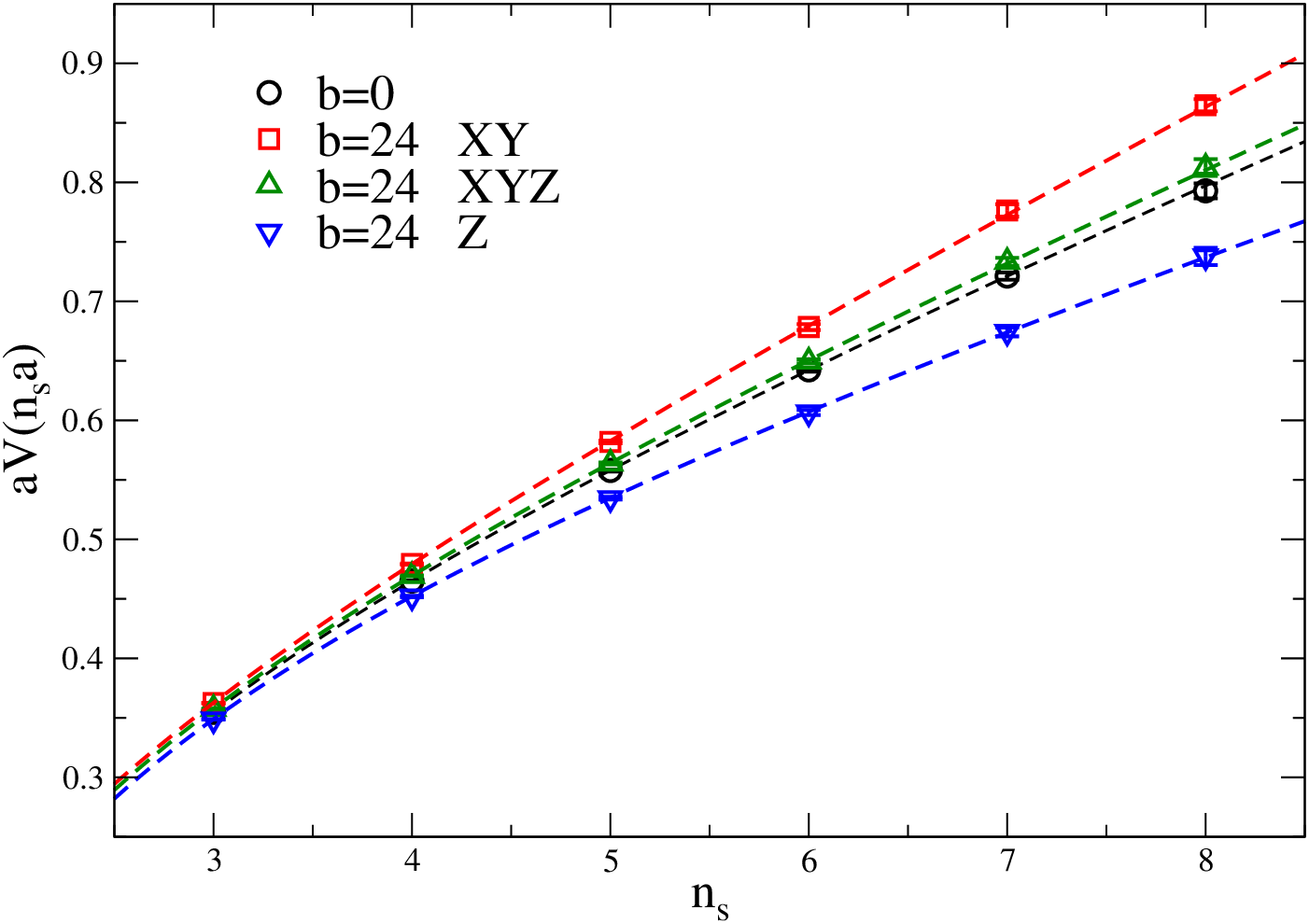}
\caption{$\overline{Q}Q-$potential both for $|e|B=0$ and for $|e|B=0.7~\textnormal{GeV}^2$ on the $40^4$ lattice.}
\label{fig_potential}
\end{figure}

In Fig. \ref{fig_potential} we report an example 
of the potential, as a function of the $\bar Q Q$ separation,
determined for our smallest lattice spacing and for two values of the 
magnetic field, $e B = 0$ and $e B \simeq 0.7$ GeV$^2$ ($b = 24$).
In the former case we averaged over all spatial directions.
For $B \neq 0$ we observe a clear anisotropic behavior, with a
striking separation of the values of the potential
measured along the $Z$ or $XY$ directions.
A comparison with $B = 0$ shows that
the potential increases in the 
transverse directions and decreases in the longitudinal direction.
This behavior is observed for all the explored lattice spacings, 
starting  from magnetic fields of the order $eB\simeq 0.2\mbox{ GeV}^2$.

In the same figure we also show, for $B \neq 0$, 
the potential obtained by averaging Wilson loops over all 
spatial directions (denoted by $XYZ$). It 
is interesting to notice that in this case the effect of 
$B$ on the static potential is strongly reduced. 
This fact may explain why previous studies have not observed significant effects of the external
field on the static potential~\cite{reg0}.

In order to characterize the dependence of the potential
on the magnetic field, we fitted it, for each value of $B$ and for 
transverse and longitudinal directions separately, 
according to the standard 
Cornell parametrization:
\begin{equation}
aV(a n \hat d )=\hat c_d +\hat{\sigma}_{d} n  + \frac{{\alpha}_d}{n},
\label{eq:potential}
\end{equation}
where $\hat d$ is a versor along the 
$z$ or $xy$ directions,
$\hat{\sigma}_d$ is the string tension, ${\alpha}_d$ the 
Coulomb coupling, and $\hat c_d$ a constant term (the index $d$
takes into account the possible dependence on the 
direction).
Such a parametrization fits reasonably well
the measured potentials, with $\chi^2/{\rm d.o.f.} \lesssim 1$
for all the explored fields and 
in a distance range going from $\sim 0.3$ to $\sim 1$ fm.
In our fits we set 
${\alpha}_d = \hat{r}_{0d}^2\hat\sigma_d -1.65$, in order 
to determine the string tension and the Sommer parameter
$\hat{r}_{0d}$ as independent quantities. A bootstrap 
analysis has been performed to determine 
the statistical errors of the 
fit parameters.

\begin{figure}[t!]
\includegraphics*[width=0.95\columnwidth]{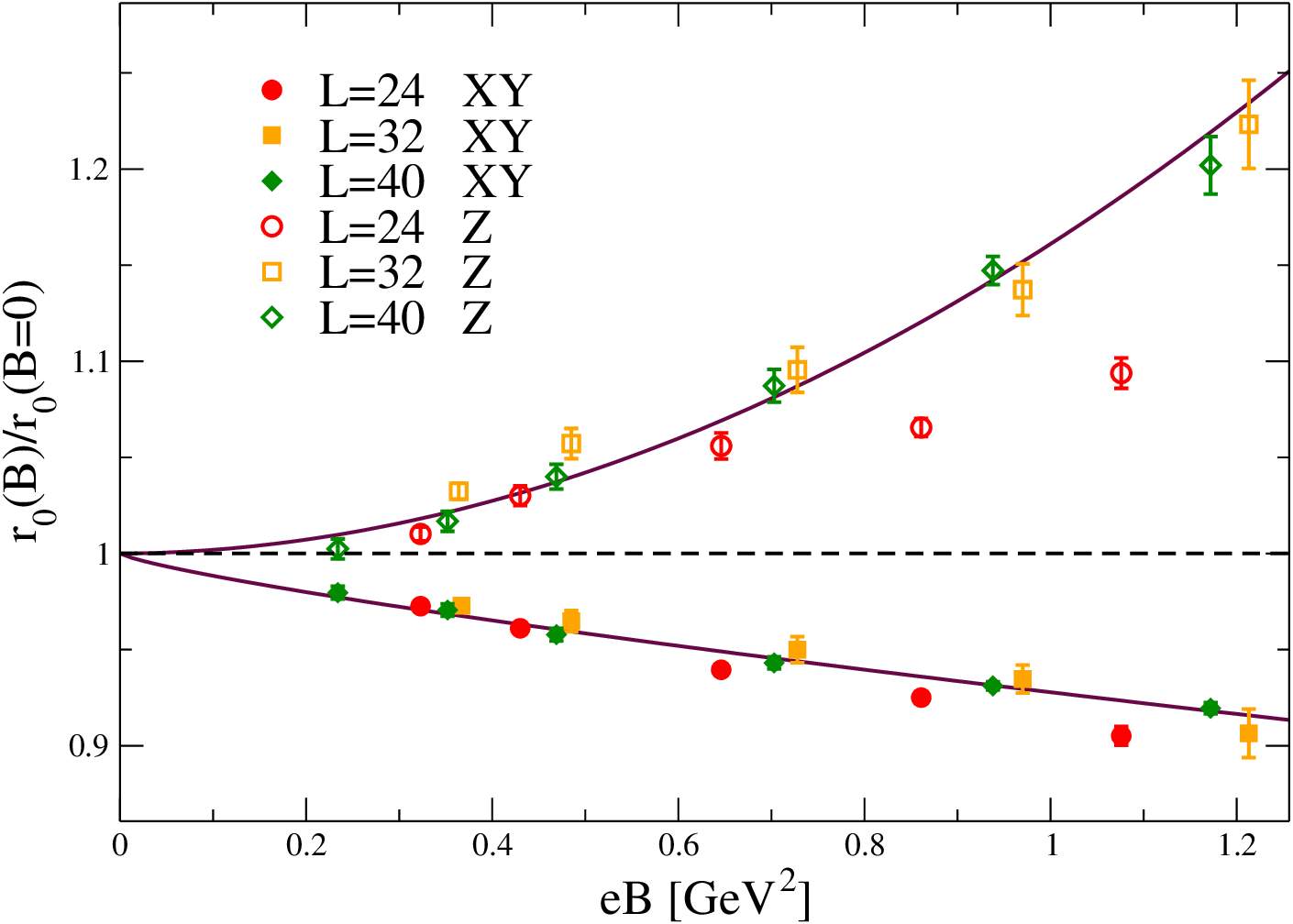}
\caption{$R^{r_0}$ along the $Z$ and $XY$ directions.
The solid line is the fit according to Eq.~(\ref{eq:ffunction}) for the $L=40$ data: 
we obtained $A^{r_{0xy}}=-0.072(2)$, $C^{r_{0xy}}=0.79(5)$ with $\chi^2/\mathrm{d.o.f.}=0.59$ 
and $A^{r_{0z}}=0.161(6)$, $C^{r_{0z}}=1.9(1)$ with $\chi^2/\mathrm{d.o.f.}=1.28$. }
\label{fig_sommer}
\end{figure}

\begin{figure}[t!]
\includegraphics*[width=0.95\columnwidth]{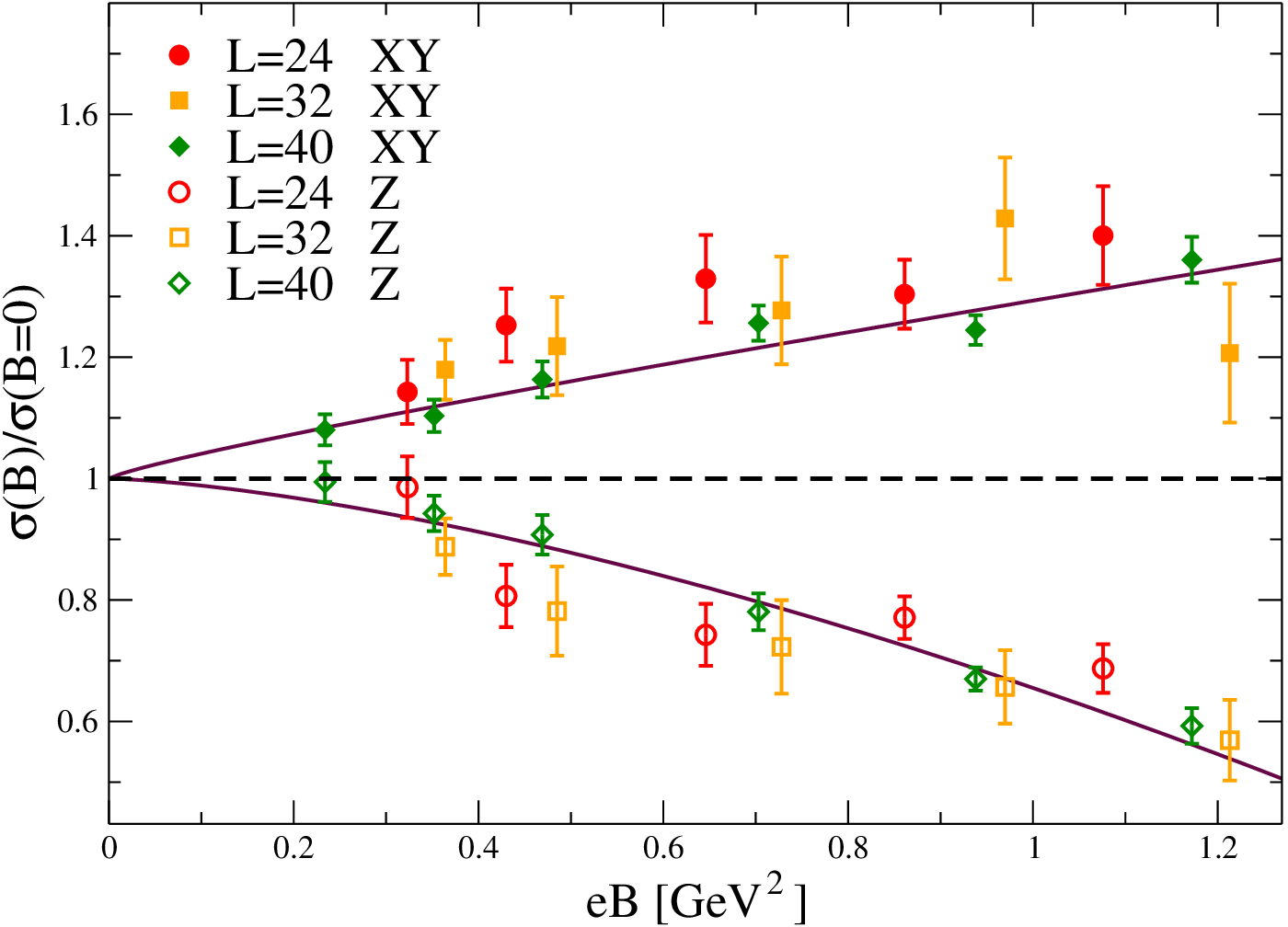}
\caption{$R^{\sigma}$ along the $Z$ and $XY$ directions. The solid line is the fit according to Eq.~(\ref{eq:ffunction}) for the $L=40$ data: we obtained $A^{\sigma_{xy}}=0.29(2)$, $C^{\sigma_{xy}}=0.9(1)$ with $\chi^2/\mathrm{d.o.f.}=1.14$ and $A^{\sigma_{z}}=-0.34(1)$, $C^{\sigma_{z}}=1.5(1)$ with $\chi^2/\mathrm{d.o.f.}=0.92$.}
\label{fig_string}
\end{figure}

\begin{figure}[t!]
\includegraphics*[width=0.95\columnwidth]{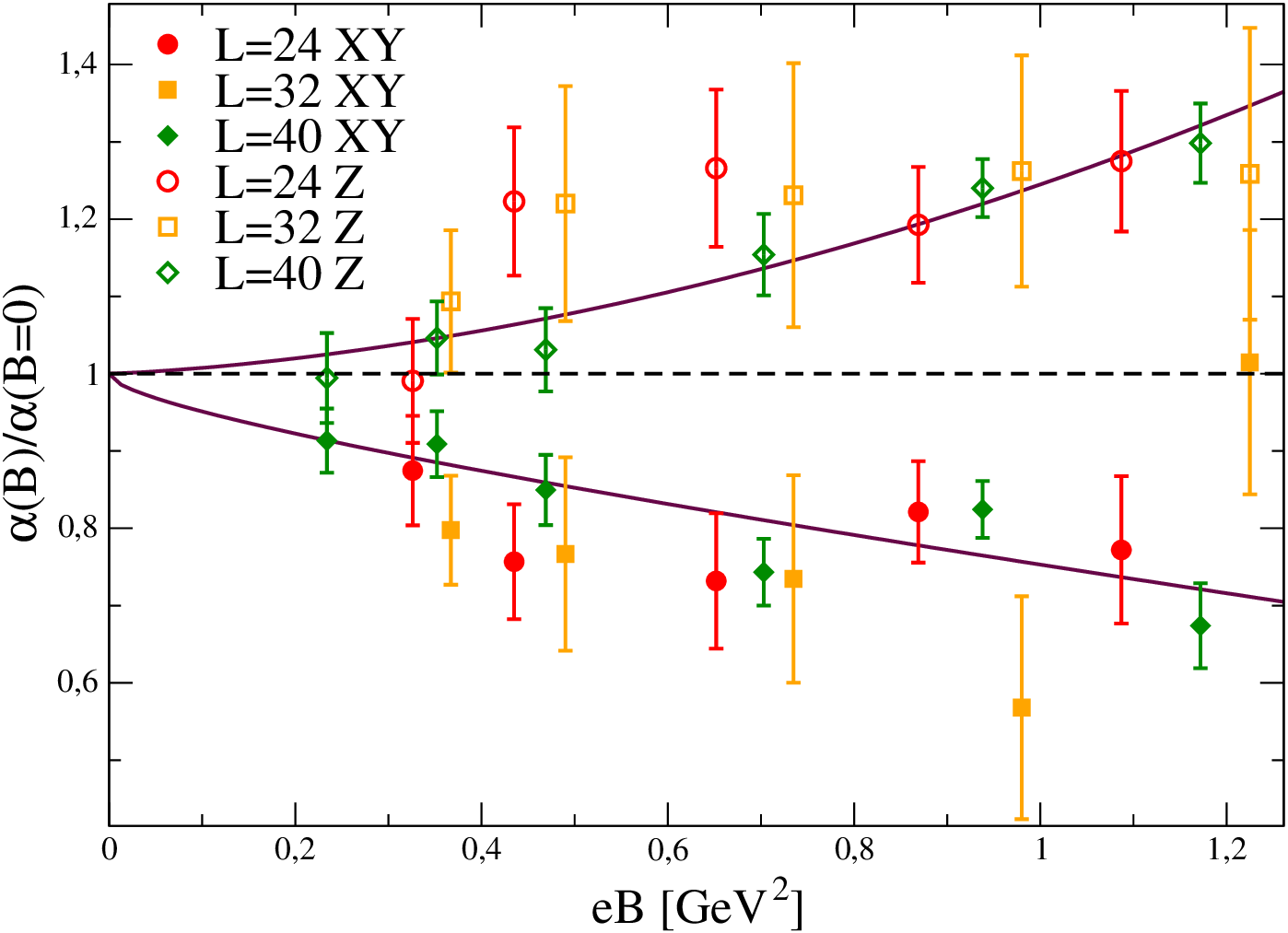}
\caption{$R^{\alpha}$ along the $Z$ and $XY$ directions. The solid line is the fit according 
to Eq.~(\ref{eq:ffunction}) for the $L=40$ data: we obtained $A^{\alpha_{xy}}=-0.24(3)$, $C^{\alpha_{xy}}=0.7(2)$ 
with $\chi^2/\mathrm{d.o.f.}=1.53$ and $A^{\alpha_{z}}=0.24(3)$, $C^{\alpha_{z}}=1.7(4)$ with $\chi^2/\mathrm{d.o.f.}=0.32$.}
\label{fig_alpha}
\end{figure}

In order to monitor the dependence of the fitted parameters on $B$,
we normalize them to the values they take for $B = 0$
at the same lattice spacing, i.e. we determine the quantities
\begin{equation}
R^{\mathcal{O}_d}=\frac{\mathcal{O}_{d}(|e|B)}{\mathcal{O}(|e|B=0)}\ ,
\end{equation}
which are shown in 
Figs.~\ref{fig_sommer}, \ref{fig_string} and \ref{fig_alpha},
respectively for 
$\mathcal{O}_d=\hat{r}_{0d}$, 
$\hat{\sigma}_d$, 
and $\alpha_d$.

The observed anisotropy in the potential 
reflects in these quantities,
leading to a significant splitting of the corresponding ratios,
which are of the order of $10-20\%$.
In particular, the string tension 
increases (decreases), as a function of $e B$, 
in the trasverse (longitudinal) direction, while the 
$\hat{r}_0$ and the Coulomb coupling show an opposite behavior.

Our results show a mild dependence on the lattice spacing,
apart from the largest fields on the coarsest lattice, for which
$e B \sim 1/a^2$; however the present accuracy of our data does not permit
a proper continuum extrapolation.
We fit the 
$L = 40$ data (corresponding to our finest lattice spacing) according to
the following ansatz for the dependence of each ratio on $B$:
\begin{equation}
R^{\mathcal{O}_d}=1+ A^{\mathcal{O}_d}\, (|e|B)^{C^{\mathcal{O}_d}} \, ;
\label{eq:ffunction}
\end{equation}
the best fit results are shown as solid lines in the figures.

\section{Discussion and conclusions}
\label{discussion}

The weak dependence of our data on $a$ suggests that what
we observed is a genuine continuum phenomenon.
However, it is important to exclude other
possibilities,
like for instance a non-trivial,
anisotropic dependence of the lattice spacing on $B$. In particular,
an increase of $a_z$ and a decrease of $a_{xy}$ as a function of $B$
would lead to similar observations.
Indeed, the magnetic field in the $\hat z$ direction
breaks the original 
$SO(4)$ symmetry of the Euclidean 
theory down to a $SO(2)_{xy}\times SO(2)_{zt}$ symmetry.
A possible effect, in the lattice regularized theory,
could be an anisotropy in the lattice spacings, which however
should still respect
$a_z = a_t$ and $a_x = a_y$.

An investigation of the dependence of the lattice spacing 
on $B$ has been reported in Ref.~\cite{reg0}, leading to the conclusion
that the dependence is not significant. Part of the evidence, which
is based on the analysis of $r_0$, is not useful,
in view of the fact that we observe an anisotropy for this 
parameter.
However, the 
dependence of the charged pion mass on $e B$, which is reported 
in Fig.~1-left of Ref.~\cite{reg0} and involves both 
$a_t$ (to get the physical value of the pion mass) 
and $a_{xy}$ (to get the physical value of $e B$)  
clearly shows that such a non-trivial
dependence is not present, at least for $eB$ up to $\sim 0.4$ GeV$^2$.
On the other hand the anisotropies 
that we observe are already 
clearly visible for such field values.

The physical origin of the observed anisotropy has to be searched
in the effective couplings between electromagnetic 
and chromoelectric/chromomagnetic fields, which stem
from quark loop contributions.
At the perturbative level~\cite{reg2}
the effective action predicts an increase of the chromoelectric
field components orthogonal to $\vec B$ (see also Ref.~\cite{galilo}), 
and a suppression of the 
longitudinal one; this is also in agreement with the observed
anisotropies at the plaquette level~\cite{Ilgenfritz:2012fw, reg2}. 
Since confinement is related to the formation of a chromoelectric 
flux-tube,
this result
suggests an increase (decrease) of the string tension in the direction 
trasverse (parallel) to $\vec B$, as we have found.
Possible anisotropies in the static potential have been predicted 
also in Ref.~\cite{anisotropic}, in particular a decrease of the 
Coulomb coupling in the transverse direction, which is consistent
with our observations.

Our exploratory study claims for extensions in various directions.
While in this context we have limited ourselves to 
the computation of the potential in the $Z$ 
and $XY$ directions, future studies will have to investigate 
its dependence on the angle $\varphi$ 
with  respect to the magnetic field direction.
On general grounds, this dependence 
is expected to be even in $\cos \varphi$ (by charge conjugation invariance).
Another interesting feature of the modifications in the potential, 
which should be better understood in the future, is the fact that they
get largely suppressed when one averages Wilson loops over all
spatial directions.
A study of the flux tube profile for the various directions 
will be necessary to get a clear picture about $B$-dependent modifications
of the chromoelectric fields.
The possibility that the string tension may vanish in the $z$ direction,
for some critical value $B$, is another interesting issue for
further investigation. 

Another direction is related to the possible
phenomenological consequences of our findings. To that aim, it will
be necessary to extend the investigation to finite temperature,
in order to have predictions valid also 
in the context of non-central heavy ion collisions.
The spectrum of mesons in the presence of a strong magnetic
background has attracted much interest in the recent
past~\cite{rhomass1, rhomass2, rhomass3, rhomass4, simonov, Machado1, Machado2, Alford, Filip},
and will likely get corrections by the presence
of anisotropies in the quark-antiquark potential.
We stress that even slight changes in the energy levels may induce
sizable corrections to cross section, {production 
and decay rates}~\cite{Machado1, Filip}. In this respect,
a direct lattice measurement of heavy quark bound states
in the presence of a magnetic background will also be of great importance.

Finally, the fact that inter-quark forces
are different in the directions
parallel or orthogonal to the magnetic field, may 
lead to anisotropies in the string breaking, as well as in  
hadronization and thermalization processes. That could have direct
consequences on various experimental probes, including
the elliptic flow.

\acknowledgments

We thank M. Chernodub and E. Laermann for useful discussions.
FS thanks M. Kruse for useful discussions regarding
code optimization on the Blue Gene/Q machine.
FN acknowledges financial support from the EU under 
project Hadron Physics 3 (Grant Agreement n. 283286).
This work was partially supported by the INFN SUMA project.
Numerical simulations have been performed on the 
Blue Gene/Q Fermi machine at CINECA, based on the 
agreement between INFN and CINECA (under INFN projects PI12 and NPQCD).


\begin{thebibliography}{99}

\bibitem{lecnotmag} 
  D.~Kharzeev, K.~Landsteiner, A.~Schmitt and H.~-U.~Yee,
  %``Strongly Interacting Matter in Magnetic Fields,''
  Lect. Notes Phys.  {\bf 871}, 1 (2013).

\bibitem{magnetars}
  R.~C.~Duncan and C.~Thompson,
  %``Formation of very strongly magnetized neutron stars - implications for
  %gamma-ray bursts,''
  Astrophys.\ J.\  {\bf 392}, L9 (1992).

\bibitem{hi1}
V.~Skokov, A.~Y.~Illarionov and V.~Toneev,
  %``Estimate of the magnetic field strength in heavy-ion collisions,''
  Int.\ J.\ Mod.\ Phys.\ A {\bf 24}, 5925 (2009)
  [arXiv:0907.1396 [nucl-th]].
  %%CITATION = ARXIV:0907.1396;%%

\bibitem{hi2}
V.~Voronyuk, V.~D.~Toneev, W.~Cassing, E.~L.~Bratkovskaya, V.~P.~Konchakovski and S.~A.~Voloshin,
  %``(Electro-)Magnetic field evolution in relativistic heavy-ion collisions,''
  Phys.\ Rev.\ C {\bf 83}, 054911 (2011)
  [arXiv:1103.4239 [nucl-th]].

\bibitem{hi3}
A.~Bzdak and V.~Skokov,
  %``Event-by-event fluctuations of magnetic and electric fields in heavy ion collisions,''
  Phys.\ Lett.\ B {\bf 710}, 171 (2012)
  [arXiv:1111.1949 [hep-ph]].

\bibitem{hi4}
W.~-T.~Deng and X.~-G.~Huang,
  %``Event-by-event generation of electromagnetic fields in heavy-ion collisions,''
  Phys.\ Rev.\ C {\bf 85}, 044907 (2012)
  [arXiv:1201.5108 [nucl-th]].
  %%CITATION = ARXIV:1201.5108;%%

\bibitem{tuchin} 
  K.~Tuchin,
  %``Particle production in strong electromagnetic fields in relativistic heavy-ion collisions,''
  Adv.\ High Energy Phys.\  {\bf 2013}, 490495 (2013)
  [arXiv:1301.0099].

\bibitem{vacha}
  T.~Vachaspati,
  %``Magnetic fields from cosmological phase transitions,''
  Phys. Lett. B {\bf 265}, 258 (1991).

\bibitem{grarub}
  D.~Grasso and H.~R.~Rubinstein,
  %``Magnetic fields in the early universe,''
  Phys. Rept.  {\bf 348}, 163 (2001)
  [astro-ph/0009061].


\bibitem{anisotropic}
  V.~A.~Miransky and I.~A.~Shovkovy,
  %``Magnetic catalysis and anisotropic confinement in QCD,''
  Phys.\ Rev.\  D \textbf{66}, 045006 (2002);

\bibitem{chernodub}
M.~N.~Chernodub,
  %``Background magnetic field stabilizes QCD string against breaking,''
  arXiv:1001.0570 [hep-ph].

\bibitem{musak}
M.~M.~Musakhanov and F.~C.~Khanna,
  %``The Axial anomaly and the conversion of gluons into photons,''
  hep-ph/9605232.

\bibitem{elze1} 
H.~T.~Elze and J.~Rafelski,
  %``Electromagnetic fields in the QCD vacuum,''
  In *Sandansky 1998, Frontier tests of QED and physics of the vacuum* 425-439
  [hep-ph/9806389].

\bibitem{elze2}
H.~T.~Elze, B.~Muller and J.~Rafelski,
  %``Interfering QCD / QED vacuum polarization,''
  hep-ph/9811372.

\bibitem{mueller} 
  M.~Asakawa, A.~Majumder and B.~Muller,
  %``Electric Charge Separation in Strong Transient Magnetic Fields,''
  Phys.\ Rev.\ C {\bf 81}, 064912 (2010).
  [arXiv:1003.2436 [hep-ph]].

\bibitem{galilo} 
  B.~V.~Galilo and S.~N.~Nedelko,
  %``Impact of the strong electromagnetic field on the QCD effective potential for homogeneous Abelian gluon field configurations,''
  Phys.\ Rev.\ D {\bf 84}, 094017 (2011).
  [arXiv:1107.4737 [hep-ph]].

\bibitem{simonov2} 
  M.~A.~Andreichikov, V.~D.~Orlovsky and Y.~.A.~Simonov,
  %``Asymptotic Freedom in Strong Magnetic Fields,''
  Phys.\ Rev.\ Lett.\  {\bf 110}, no. 16, 162002 (2013)
  [arXiv:1211.6568 [hep-ph]].
  %%CITATION = ARXIV:1211.6568;%%



\bibitem{KojoSu1} 
  T.~Kojo and N.~Su,
  %``The quark mass gap in a magnetic field,''
  Phys.\ Lett.\ B {\bf 720}, 192 (2013)
  [arXiv:1211.7318 [hep-ph]].

\bibitem{KojoSu2} 
  T.~Kojo and N.~Su,
  %``A renormalization group approach for QCD in a strong magnetic field,''
  Phys.\ Lett.\ B {\bf 726}, 839 (2013)
  [arXiv:1305.4510 [hep-ph]].

\bibitem{watson} 
  P.~Watson and H.~Reinhardt,
  %``Quark gap equation in an external magnetic field,''
  Phys.\ Rev.\ D {\bf 89}, 045008 (2014)
  [arXiv:1310.6050 [hep-ph]].

\bibitem{andersen} 
  J.~O.~Andersen, W.~R.~Naylor and A.~Tranberg,
  %``Chiral and deconfinement transitions in a magnetic background using the functional renormalization group with the Polyakov loop,''
  arXiv:1311.2093 [hep-ph].

\bibitem{ozaki}
S.~Ozaki,
  %``QCD effective potential with strong U(1)_{em} magnetic fields,''
  Phys.\ Rev.\ D \textbf{89}, 054022 (2014)
  [arXiv:1311.3137 [hep-ph]].

\bibitem{kamikado} 
  K.~Kamikado and T.~Kanazawa,
  %``Chiral dynamics in a magnetic field from the functional renormalization group,''
  JHEP {\bf 1403}, 009 (2014)
  [arXiv:1312.3124 [hep-ph]].

\bibitem{mueller2} 
  N.~Mueller, J.~A.~Bonnet and C.~S.~Fischer,
  %``Dynamical quark mass generation in a strong external magnetic field,''
  arXiv:1401.1647 [hep-ph].



\bibitem{demusa}
  M.~D'Elia, S.~Mukherjee and F.~Sanfilippo,
  %``QCD Phase Transition in a Strong Magnetic Background,''
  Phys.\ Rev.\ D \textbf{82}, 051501 (2010)
  [arXiv:1005.5365 [hep-lat]].

\bibitem{DEN}
  M.~D'Elia and F.~Negro, 
  %``Chiral Properties of Strong Interactions in a Magnetic Background''
  Phys. Rev. D {\bf 83}, 114028 (2011)
  [arXiv:1103.2080 [hep-lat]].

\bibitem{reg0}
  G.~S.~Bali, F.~Bruckmann, G.~Endrodi, Z.~Fodor, S.~D.~Katz, S.~Krieg, A.~Schafer and K.~K.~Szabo,
  %``The QCD phase diagram for external magnetic fields''
  JHEP \textbf{1202}, 044 (2012)
  [arXiv:1111.4956 [hep-lat]].

\bibitem{Ilgenfritz:2012fw} 
  E.~-M.~Ilgenfritz, M.~Kalinowski, M.~Muller-Preussker, B.~Petersson and A.~Schreiber,
  %``Two-color QCD with staggered fermions at finite temperature under the influence of a magnetic field,''
  Phys.\ Rev.\ D {\bf 85}, 114504 (2012)
  [arXiv:1203.3360 [hep-lat]].

\bibitem{reg2}
G.~S.~Bali, F.~Bruckmann, G.~Endrodi, F.~Gruber and A.~Schaefer,
  %``Magnetic field-induced gluonic (inverse) catalysis and pressure (an)isotropy in QCD,''
  JHEP \textbf{1304}, 130 (2013)
  [arXiv:1303.1328 [hep-lat]].

\bibitem{EB} 
  M.~D'Elia, M.~Mariti and F.~Negro,
  %``Susceptibility of the QCD vacuum to CP-odd electromagnetic background fields,''
  Phys.\ Rev.\ Lett.\  {\bf 110}, 082002 (2013)
  [arXiv:1209.0722 [hep-lat]].


\bibitem{kovacs}
  F.~Bruckmann, G.~Endrodi and T.~G.~Kovacs,
  %``Inverse magnetic catalysis and the Polyakov loop,''
  JHEP {\bf 1304}, 112 (2013)
  [arXiv:1303.3972 [hep-lat]].

\bibitem{Ilgenfritz:2013ara} 
  E.~-M.~Ilgenfritz, M.~Muller-Preussker, B.~Petersson and A.~Schreiber,
  %``Magnetic catalysis (and inverse catalysis) at finite temperature in two-color lattice QCD,''
  arXiv:1310.7876 [hep-lat].

\bibitem{catalreview}
I.~A.~Shovkovy,
  %``Magnetic Catalysis: A Review,''
  Lect.\ Notes Phys.\  {\bf 871}, 13 (2013)
  [arXiv:1207.5081 [hep-ph]].
  %%CITATION = ARXIV:1207.5081;%%

\bibitem{fukuhida} 
  K.~Fukushima and Y.~Hidaka,
  %``Magnetic Catalysis vs Magnetic Inhibition,''
  Phys.\ Rev.\ Lett.\  {\bf 110}, 031601 (2013)
  [arXiv:1209.1319 [hep-ph]].
  %%CITATION = ARXIV:1209.1319;%%

\bibitem{Bornyakov:2013eya} 
  V.~G.~Bornyakov, P.~V.~Buividovich, N.~Cundy, O.~A.~Kochetkov and A.~Schäfer,
  %``Deconfinement transition in two-flavour lattice QCD with dynamical overlap fermions in an external magnetic field,''
  arXiv:1312.5628 [hep-lat].

\bibitem{Chao:2013qpa} 
  J.~Chao, P.~Chu and M.~Huang,
  %``Inverse magnetic catalysis induced by sphalerons,''
  Phys.\ Rev.\ D {\bf 88}, 054009 (2013)
  [arXiv:1305.1100 [hep-ph]].

\bibitem{Fraga:2013ova} 
  E.~S.~Fraga, B.~W.~Mintz and J.~Schaffner-Bielich,
  %``A search for inverse magnetic catalysis in thermal quark-meson models,''
  Phys.\ Lett.\ B {\bf 731}, 154 (2014)
  [arXiv:1311.3964 [hep-ph]].

\bibitem{Yu:2014sla} 
  L.~Yu, H.~Liu and M.~Huang,
  %``Spontaneous generation of local CP violation and inverse magnetic catalysis,''
  arXiv:1404.6969 [hep-ph].

\bibitem{Ferreira:2014kpa} 
  M.~Ferreira, P.~Costa, O.~Lourenço, T.~Frederico and C.~Providência,
  %``Magnetic inverse catalysis in the (2+1)-flavor Nambu--Jona-Lasinio and Polyakov--Nambu--Jona-Lasinio models,''
  arXiv:1404.5577 [hep-ph].

\bibitem{Farias:2014eca} 
  R.~L.~S.~Farias, K.~P.~Gomes, G.~I.~Krein and M.~B.~Pinto,
  %``The Importance of Asymptotic Freedom for the Pseudocritical Temperature in Magnetized Quark Matter,''
  arXiv:1404.3931 [hep-ph].

\bibitem{Ruggieri:2014bqa} 
  M.~Ruggieri, L.~Oliva, P.~Castorina, R.~Gatto and V.~Greco,
  %``Critical Endpoint and Inverse Magnetic Catalysis for Finite Temperature and Density Quark Matter in a Magnetic Background,''
  arXiv:1402.0737 [hep-ph].


\bibitem{tcwup1} 
  Y.~Aoki, S.~Borsanyi, S.~Durr, Z.~Fodor, S.~D.~Katz, S.~Krieg and K.~K.~Szabo,
  %``The QCD transition temperature: results with physical masses in the continuum limit II.,''
  JHEP {\bf 0906}, 088 (2009)
  [arXiv:0903.4155 [hep-lat]].

\bibitem{befjkkrs}
S.~Borsanyi, G.~Endrodi, Z.~Fodor, A.~Jakovac, S.~D.~Katz, S.~Krieg, C.~Ratti and K.~K.~Szabo,
  %``The QCD equation of state with dynamical quarks,''
  JHEP {\bf 1011}, 077 (2010)
  [arXiv:1007.2580 [hep-lat]].

\bibitem{weisz} 
  P.~Weisz,
  %``Continuum Limit Improved Lattice Action for Pure Yang-Mills Theory. 1.,''
  Nucl.\ Phys.\ B {\bf 212}, 1 (1983).
  %%CITATION = NUPHA,B212,1;%%
  %286 citations counted in INSPIRE as of 25 Oct 2013

\bibitem{curci} 
G.~Curci, P.~Menotti and G.~Paffuti,
  %``Symanzik's Improved Lagrangian for Lattice Gauge Theory,''
  Phys.\ Lett.\ B {\bf 130}, 205 (1983)
  [Erratum-ibid.\ B {\bf 135}, 516 (1984)].
  %%CITATION = PHLTA,B130,205;%%

\bibitem{morning} 
  C.~Morningstar and M.~J.~Peardon,
  %``Analytic smearing of SU(3) link variables in lattice QCD,''
  Phys.\ Rev.\ D {\bf 69}, 054501 (2004)
  [hep-lat/0311018].

\bibitem{bazavov} 
  A.~Bazavov {\it et al.}  [HotQCD Collaboration],
  %``Taste symmetry and QCD thermodynamics with improved staggered fermions,''
  PoS LATTICE {\bf 2010}, 169 (2010)
  [arXiv:1012.1257 [hep-lat]].

\bibitem{thooft}
G.~'t Hooft,
  %``A Property of Electric and Magnetic Flux in Nonabelian Gauge Theories,''
  Nucl. Phys. B \textbf{153}, 141 (1979).

\bibitem{bound3} 
  P.~H.~Damgaard and U.~M.~Heller,
  %``The U(1) Higgs Model In An External Electromagnetic Field,''
  Nucl. Phys. B \textbf{309}, 625 (1988).


\bibitem{wiese}
  M.~H.~Al-Hashimi and U.~J.~Wiese,
  %``Discrete Accidental Symmetry for a Particle in a Constant Magnetic Field on
  %a Torus,''
  Ann. Phys. \textbf{324}, 343 (2009)
  [arXiv:0807.0630 [quant-ph]].

\bibitem{review} 
M.~D'Elia,
  %``Lattice QCD Simulations in External Background Fields,''
  Lect.\ Notes Phys.\  {\bf 871}, 181 (2013)
  [arXiv:1209.0374 [hep-lat]].

\bibitem{Hasenfratz:2001hp} 
  A.~Hasenfratz and F.~Knechtli,
  %``Flavor symmetry and the static potential with hypercubic blocking,''
  Phys.\ Rev.\ D {\bf 64}, 034504 (2001)
  [hep-lat/0103029].

\bibitem{Della Morte:2005yc} 
  M.~Della Morte, A.~Shindler and R.~Sommer,
  %``On lattice actions for static quarks,''
  JHEP {\bf 0508}, 051 (2005)
  [hep-lat/0506008].

\bibitem{Albanese:1987ds}
 M.~Albanese {\it et al.}  [APE Collaboration],
 %``Glueball Masses and String Tension in Lattice QCD,''
 Phys.\ Lett.\ B {\bf 192} (1987) 163.

\bibitem{rhomass1}
M.~N.~Chernodub,
  %``Spontaneous electromagnetic superconductivity of vacuum in strong magnetic field: evidence from the Nambu--Jona-Lasinio model,''
  Phys.\ Rev.\ Lett.\  {\bf 106}, 142003 (2011)
  [arXiv:1101.0117 [hep-ph]].
  %%CITATION = ARXIV:1101.0117;%%
  %80 citations counted in INSPIRE as of 20 Mar 2014

\bibitem{rhomass2}
Y.~Hidaka and A.~Yamamoto,
  %``Charged vector mesons in a strong magnetic field,''
  Phys.\ Rev.\ D {\bf 87}, 094502 (2013)
  [arXiv:1209.0007 [hep-ph]].
  %%CITATION = ARXIV:1209.0007;%%

\bibitem{rhomass3} 
  M.~Frasca,
  %``$\rho$ condensation and physical parameters,''
  JHEP {\bf 1311}, 099 (2013)
  [arXiv:1309.3966 [hep-ph]].

\bibitem{rhomass4} 
  N.~Callebaut, D.~Dudal and H.~Verschelde,
  %``Holographic rho mesons in an external magnetic field,''
  JHEP {\bf 1303}, 033 (2013)
  [arXiv:1105.2217 [hep-th]].

\bibitem{simonov} 
  M.~A.~Andreichikov, B.~O.~Kerbikov, V.~D.~Orlovsky and Y.~A.~Simonov,
  %``Meson Spectrum in Strong Magnetic Fields,''
  Phys.\ Rev.\ D {\bf 87}, 094029 (2013)
  [arXiv:1304.2533 [hep-ph]].

\bibitem{Machado1} 
  C.~S.~Machado, F.~S.~Navarra, E.~G.~de Oliveira, J.~Noronha and M.~Strickland,
  %``Heavy quarkonium production in a strong magnetic field,''
  Phys.\ Rev.\ D {\bf 88}, 034009 (2013)
  [arXiv:1305.3308 [hep-ph]].

\bibitem{Machado2} 
  C.~S.~Machado, S.~I.~Finazzo, R.~D.~Matheus and J.~Noronha,
  %``Modification of the B Meson Mass in a Magnetic Field from QCD Sum Rules,''
  arXiv:1307.1797.

\bibitem{Alford} 
  J.~Alford and M.~Strickland,
  %``Charmonia and Bottomonia in a Magnetic Field,''
  Phys.\ Rev.\ D {\bf 88}, 105017 (2013)
  [arXiv:1309.3003 [hep-ph]].

\bibitem{Filip} 
  P.~Filip,
  %``Heavy Flavor Mesons in Strong Magnetic Fields,''
  PoS CPOD {\bf 2013}, 035 (2013).



\end{thebibliography}
\end{document}